\documentclass[12pt]{iopart}
\usepackage{iopams}
\usepackage{graphicx}

\begin{document}

\title[Thermal expansion and pressure in Cu$_x$TiSe$_2$]
{Thermal expansion and effect of pressure on superconductivity in Cu$_x$TiSe$_2$}

\author{S. L. Bud'ko, P. C. Canfield}
\address{Ames Laboratory US DOE and Department of Physics and Astronomy,
Iowa State University, Ames, IA 50011, USA}

\author{E. Morosan, R. J. Cava}
\address{Department of Chemistry, Princeton University, Princeton, NJ 08540, USA}

\author{G. M. Schmiedeshoff}
\address{Department of Physics, Occidental College, Los Angeles, CA 90041, USA}

\begin{abstract}

We report measurements of thermal expansion on a number of polycrystalline Cu$_x$TiSe$_2$ samples corresponding to
the parts of $x - T$ phase diagram with different ground states, as well as the pressure dependence of the
superconducting transition temperature for samples with three different values of Cu-doping. Thermal expansion
data suggest that the $x - T$ phase diagram may be more complex than initially reported. $T_c$ data at elevated
pressure can be scaled to the ambient pressure Cu$_x$TiSe$_2$ phase diagram, however, significantly different
scaling factors are needed to accommodate the literature data on the charge density wave transition suppression
under pressure.

\end{abstract}

\pacs{74.62.Fj, 65.40.De, 71.45.Lr}

\submitto{\JPCM}

\maketitle

\section{Introduction}

The transition metal dichalcogenides and their intercalate complexes received a lot of attention in the past
several decades \cite{wil69a,wil75a,fri87a} due to their low dimensionality, the tunability of their properties,
and an abundance of curious physical phenomena associated with this class of materials. Of those, TiSe$_2$ was one
of the first compounds where a charge-density-wave (CDW) transition was observed, yet, the physical mechanism
governing this transition is abstruse and the number of studies related to this material continues to grow.
Recently a new development in transition metal dichalcogenides was reported: Cu-intercalation in Cu$_x$TiSe$_2$
caused continuous suppression of the CDW transition followed by (or, initially, coexistent with) a superconducting
state near $x = 0.04$, with a maximum superconducting temperature $T_c \approx 4.15$ K for Cu$_{0.08}$TiSe$_2$.
\cite{mor06a} The physics behind the intriguing phase diagram for Cu$_x$TiSe$_2$ presented in \cite{mor06a} is
still not fully understood. In search of clues, in this work we report measurements of thermal expansion for
number of Cu-concentrations corresponding to different parts of the phase diagram, as well as the pressure
dependence of the superconducting transition temperature for three different values of Cu-doping.

\section{Experimental methods}

Polycrystalline Cu$_x$TiSe$_2$ samples were synthesized by two-step solid state reaction (see Ref \cite{mor06a}
for more details) and were in the form of homogenous purple-grey pellets which were 75\%$\pm$10\% of theoretical
density. Thermal expansion (TE) was measured approximately along the axis of the pellet, the samples were shaped
using dry diamond impregnated wire saw followed by a light, dry, sand paper polishing. For pure TiSe$_2$ thermal
expansion was measured both along the axis of the pellet and perpendicular to the axis to address the possible
preferential orientation of the grains forming the pellet. Thermal expansion was measured using a capacitive
dilatometer constructed of OFHC copper; a detailed description of the dilatometer is presented elsewhere
\cite{sch06a}. The dilatometer was mounted in a Quantum Design PPMS-14 instrument and was operated over a
temperature range of 1.8 to 300 K. The same set-up was used in our recent work on YNi$_2$B$_2$C and
ErNi$_2$B$_2$C. \cite{bud06a,bud06b}.

The piston-cylinder clamp-type pressure cell made out of non-magnetic Ni-Co alloy {\it MP35N} used in this work
was designed to fit a commercial Quantum Design MPMS-5 SQUID magnetometer (see \cite{bud05a} for a detailed
description of the cell). Pressure was generated in a teflon capsule filled with approximately 50:50 mixture of
n-pentane and mineral oil. The shift in the superconducting transition temperature of 6N purity Pb, placed in the
capsule together with the sample, was used to determine pressure at low temperatures \cite{eil81a}. DC
magnetization measurements were performed in an applied field of 25 Oe in a zero-field-cooled warming protocol.

\section{Results and discussion}

\subsection{Thermal expansion}

\begin{figure}[tbp]
\begin{center}
\includegraphics[angle=0,width=120mm]{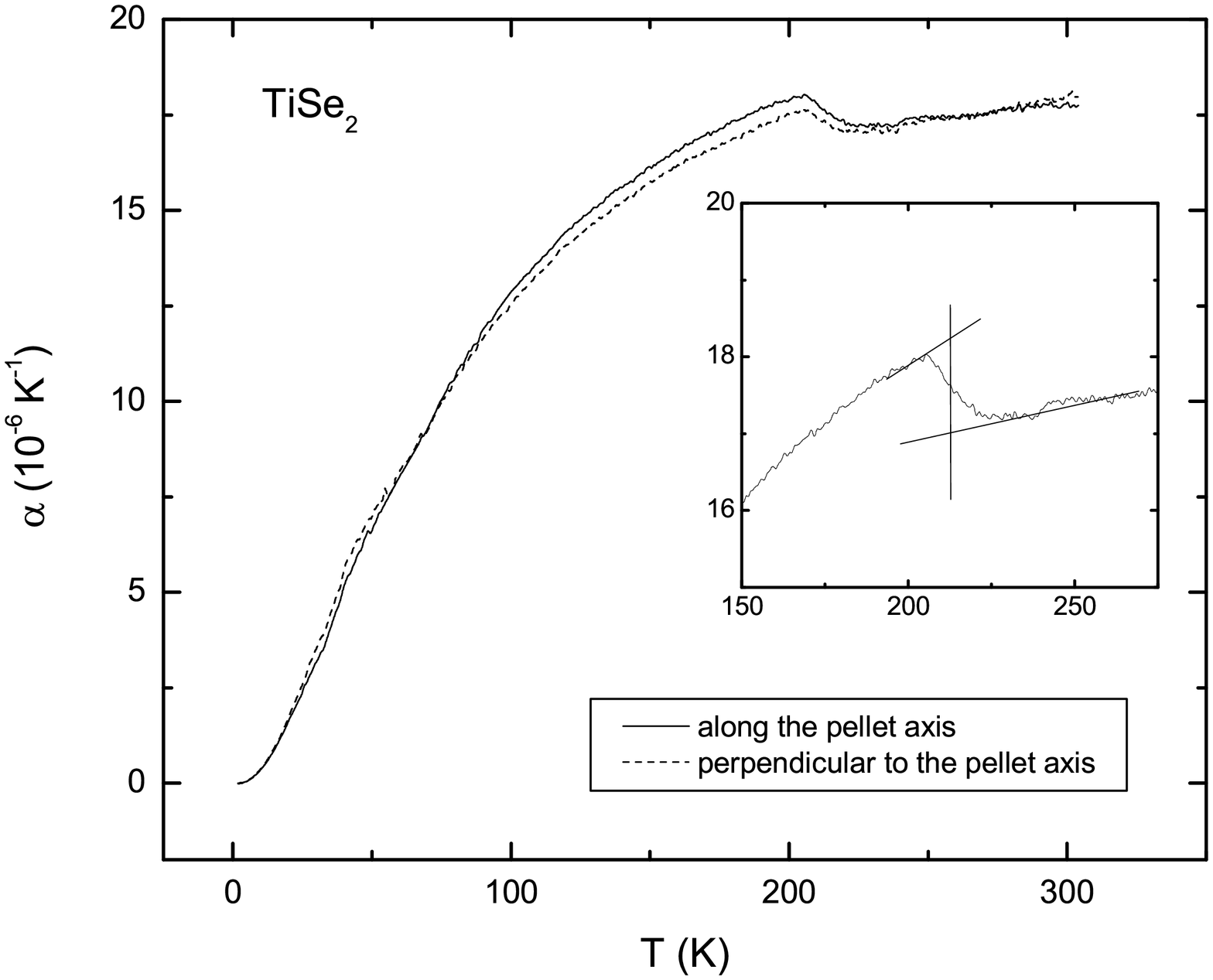}
\end{center}
\caption{Temperature-dependent thermal expansion of pure TiSe$_2$ sample measured along and perpendicular to the
pellet axis. Inset: enlarged region near $T_{CDW}$ with definitions of $T_{CDW}$ and $\Delta \alpha$ used
throughout this paper.} \label{F1}
\end{figure}

Fig. \ref{F1} shows the temperature-dependent thermal expansion coefficient of pure, polycrystalline TiSe$_2$
measured along the pellet axis as well as perpendicular to it. The two curves are very similar, suggesting that
the distribution of the grains within the sample is rather uniform. A sharp, distinct, feature in the temperature
dependence of the thermal expansion coefficient, $\alpha(T)$ at $\approx 213$ K  marks the CDW transition. The
change of $\alpha$ at the transition, defined as sketched in the inset to Fig. \ref{F1}, is, as averaged from the
two measurements, $\Delta \alpha \approx -1.2 \cdot 10^{-6}$ K$^{-1}$ (here we will use the following sign
convention: $\Delta \alpha > 0$ if $\alpha(T)$ \textit{increases} at $T_{CDW}$ on warming and vise versa).
Literature data on thermal expansion of TiSe$_2$ at $T_{CDW}$ are somewhat inconsistent and make comparison with
our data ambiguous: Weigers \cite{wei80a} reported $\Delta \alpha_c \approx 7.5 \cdot 10^{-6}$ K$^{-1}$, and no
measurable change in $\Delta \alpha_a$, whereas Caill\'{e} et al. \cite{cai83a} claimed a clear change in $a$-axis
thermal expansion coefficient at $T_{CDW}$, $\Delta \alpha_a \approx -2.5 \cdot 10^{-6}$ K$^{-1}$. From the data
of Weigers \cite{wei80a} $\Delta \alpha_{poly} = (2 \cdot \Delta \alpha_a + \Delta \alpha_c)/3 \approx \Delta
\alpha_c/3 = 2.5 \cdot 10^{-6}$ K$^{-1}$, whereas using $\Delta \alpha_c$ and $\Delta \alpha_a$ from Refs.
\cite{wei80a} and \cite{cai83a} respectively, $\Delta \alpha_{poly} \approx 0.8 \cdot 10^{-6}$ K$^{-1}$. Both of
these estimates of $\Delta \alpha_{poly}$ differ in sign and value from our measurements. The reason for this
discrepancy is not understood, however, if we take the difference in the literature values of $\Delta \alpha_a$
\cite{wei80a,cai83a} as a measure of the error bars in the literature data, this discrepancy will be lifted,
additionally, the estimate of the the thermal expansion of a polycrystal as a simple average over all directions
may be an oversimplification for an anisotropic material like TiSe$_2$ (see brief discussion in Chapter 7 of Ref.
\cite{bar99a} and references therein).

\begin{figure}[tbp]
\begin{center}
\includegraphics[angle=0,width=80mm]{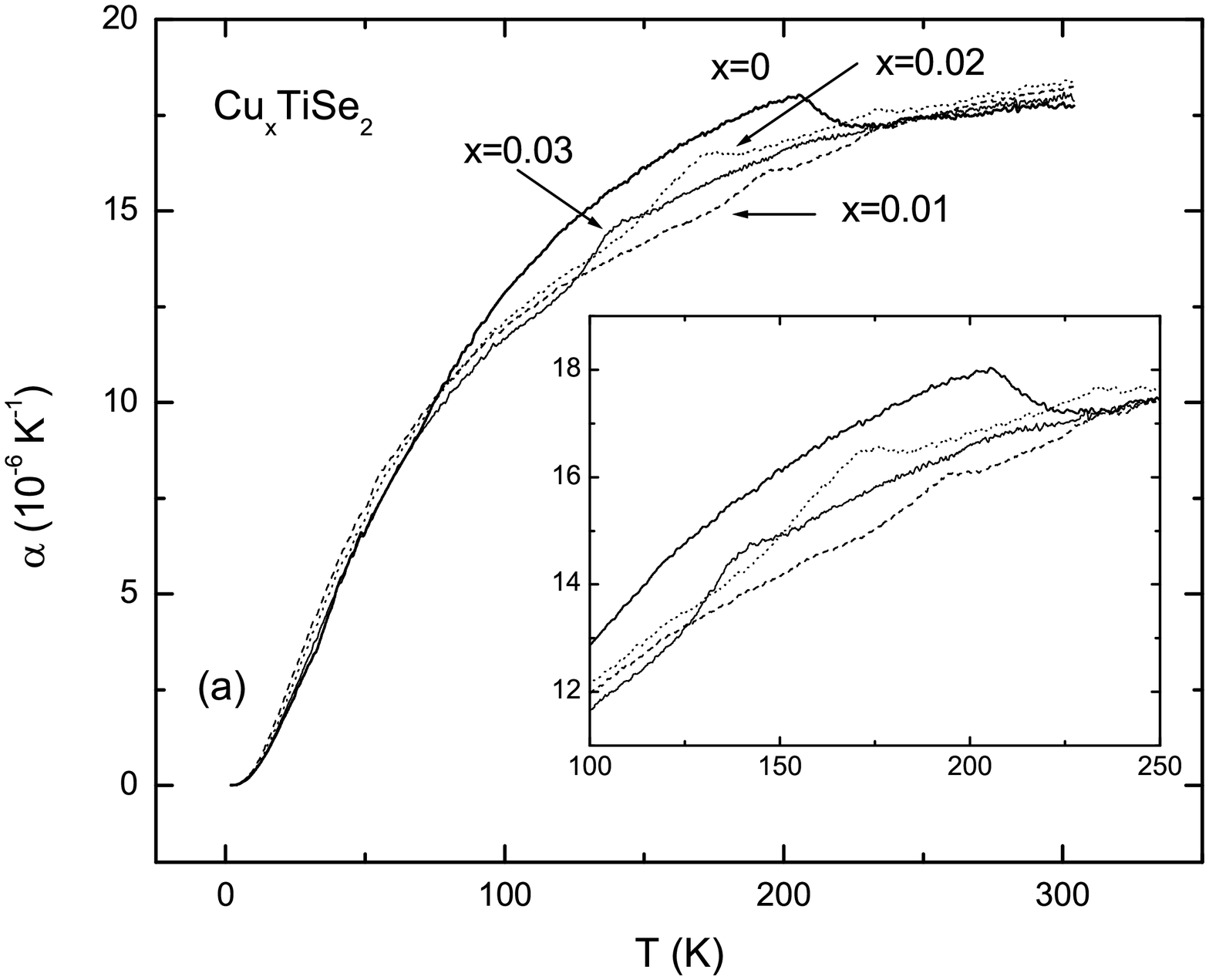}
\includegraphics[angle=0,width=80mm]{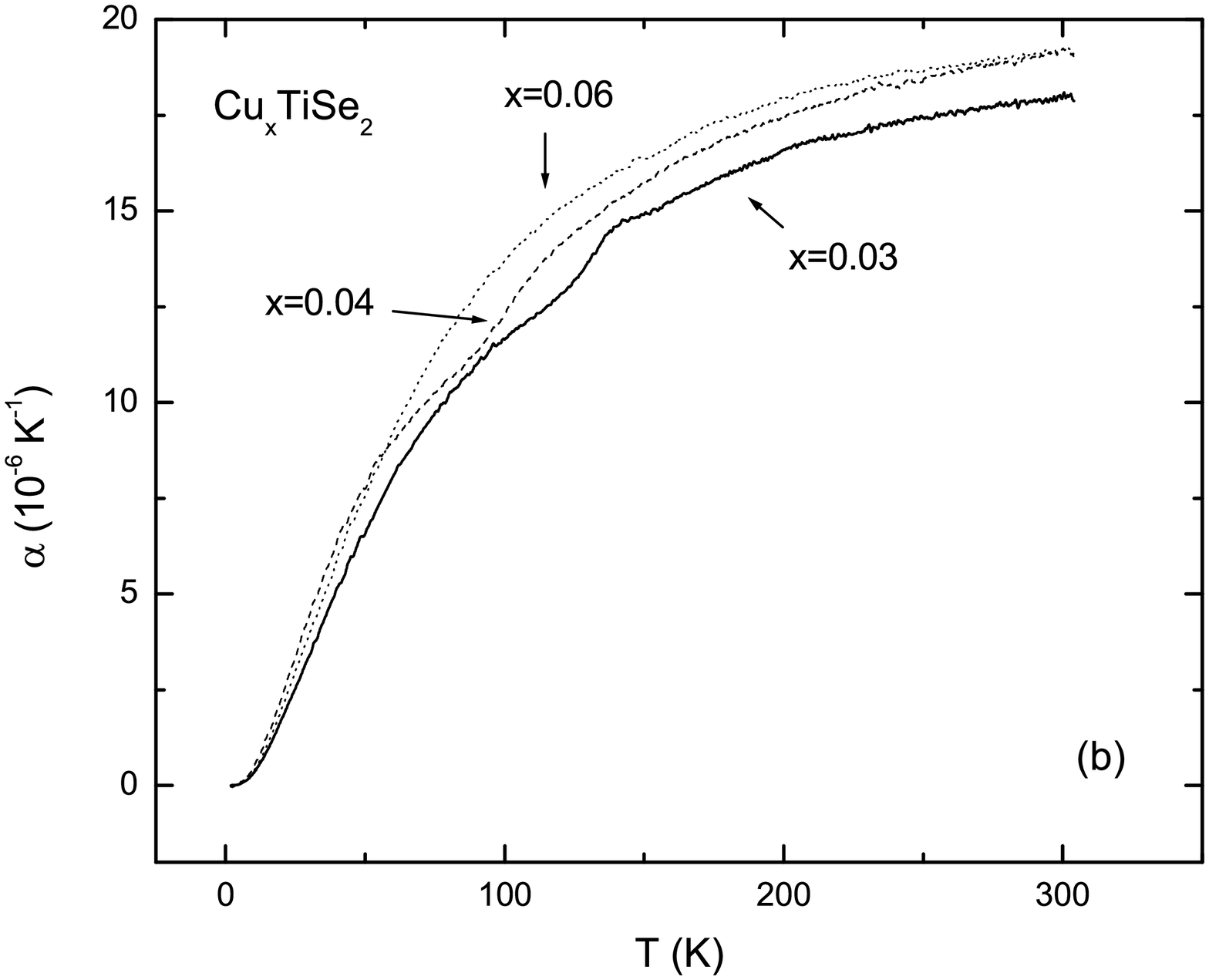}
\includegraphics[angle=0,width=80mm]{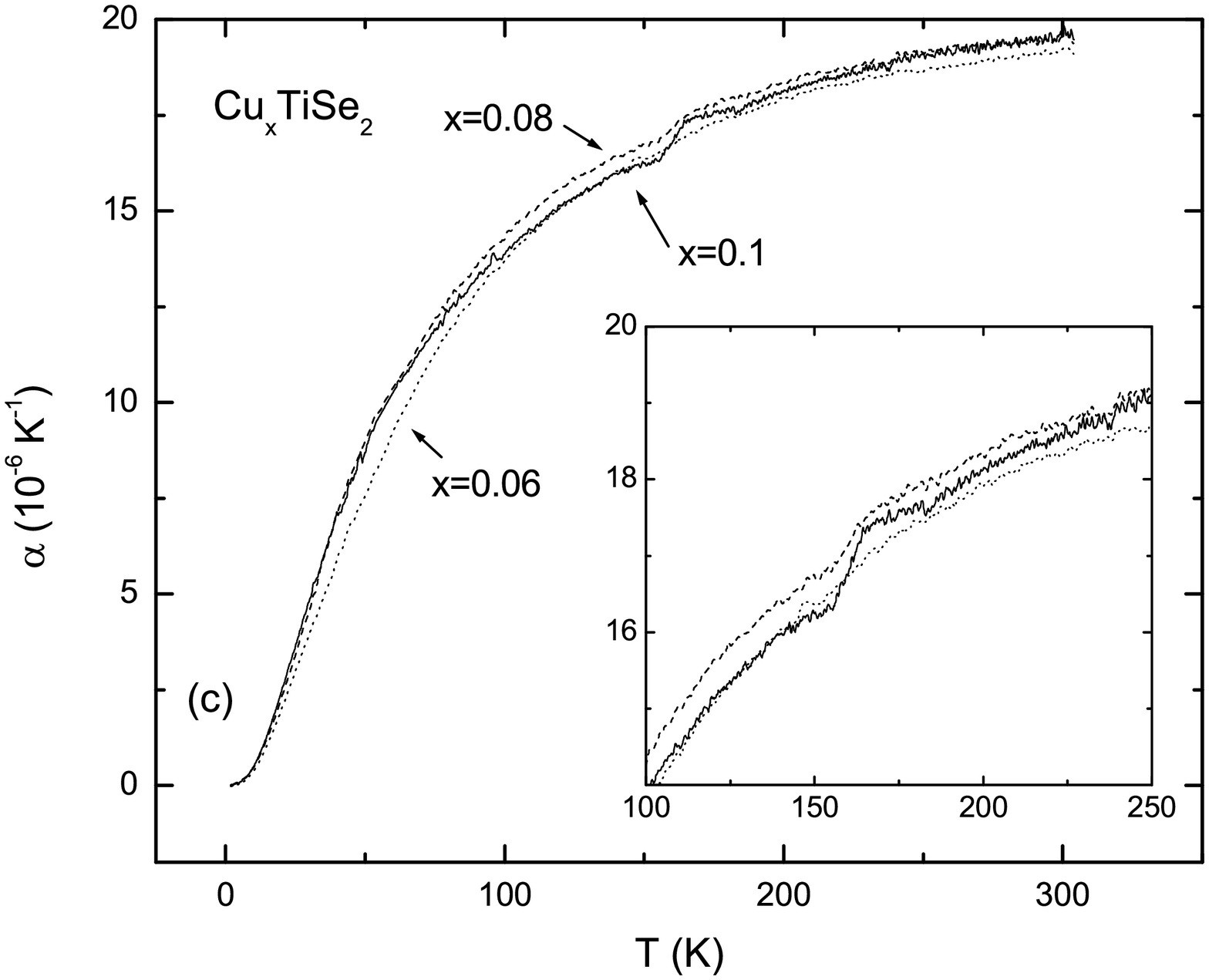}
\end{center}
\caption{Temperature-dependent thermal expansion of Cu$_x$TiSe$_2$ polycrystalline samples. Note, the data for
Cu$_{0.03}$TiSe$_2$ and Cu$_{0.06}$TiSe$_2$ are plotted twice, on different panels, for comparison. Insets to (a)
and (c): enlarged region near the features in $\alpha(T)$.} \label{F2}
\end{figure}

Temperature dependent thermal expansion coefficient for different Cu$_x$TiSe$_2$ samples is plotted in Fig.
\ref{F2}. These curves have several features of note.

\begin{figure}[tbp]
\begin{center}
\includegraphics[angle=0,width=120mm]{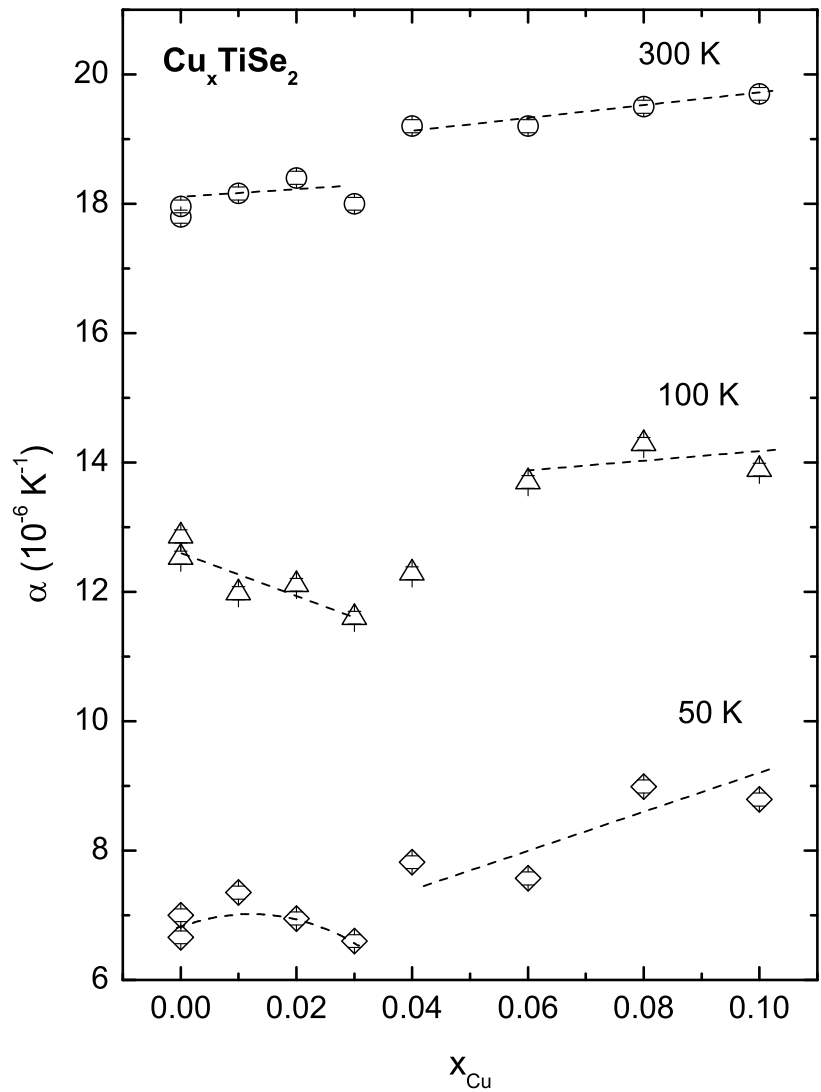}
\end{center}
\caption{Thermal expansion coefficient of polycrystalline Cu$_x$TiSe$_2$ at 300 K, 100 K and 50 K. Lines are
guides for the eye. Two points for $x = 0$ are from two measurements in Fig. \ref{F1}. Error bars are roughly
estimated from the noise in $\alpha(T)$ data near 300 K.} \label{F3}
\end{figure}

(i)For the samples in the range $0 \leq x \leq 0.03$ the thermal expansion coefficient is quite similar near room
temperature (240 K$ \leq T \leq$ 300 K) and below approximately 70 K (Fig. \ref{F2}(a)). On further increase of Cu
intercalation, between $x = 0.03$ and $x = 0.06$ (Fig. \ref{F2}(b)), $\alpha(T)$ increases in the whole
temperature range, and then the general behavior becomes very similar for $0.06 \leq x \leq 0.1$ (Fig.
\ref{F2}(c)). $\alpha(300$K$)$ plotted as a function of $x$, Cu concentration, (Fig. \ref{F3}) shows an abrupt
change between $x = 0.03$ and 0.04.

(ii)Temperature dependent thermal expansion coefficient for Cu$_x$TiSe$_2$ ($0 \leq x \leq 0.04$) samples (Fig.
\ref{F2}(a,b)) shows a clear feature at temperatures close to the $T_{CDW}$ determined from resistivity or/and
susceptibility measurements \cite{mor06a}. While for $x = 0$ $\Delta \alpha$ at CDW transition is negative, it is
positive for $x = 0.03, 0.04$ and the feature has some intermediate shape for $x = 0.01, 0.02$.

(iii)For $x = 0.08, 0.1$ a step-like feature is seen at $T \approx 160$ K. No feature in this temperature range
was reported for Cu$_{0.08}$TiSe$_2$ and Cu$_{0.1}$TiSe$_2$ samples in the previous study \cite{mor06a}.

It should be mentioned that we cannot detect superconducting transitions in our thermal expansion measurements
(for $x = 0.06, 0.08, 0.1$ $T_c$ was reported \cite{mor06a} to be above our base temperature), this is not
surprising, bearing in mind the thermodynamic Ehrenfest relations and the small values of $\Delta C_P$ at $T_c$
\cite{mor06a} and pressure derivatives $dT_c/dP$ (see below).

The features in TE for the samples with Cu concentration in the range $0 \leq x \leq 0.04$ apparently correspond
to the CDW transition, with slight differences in the characteristic temperatures probably being due to the width
of the features observed in different measurements and adopted criteria for determining of $T_{CDW}$ (see Fig.
\ref{F8} below). The rather sharp change in $\alpha_{300K}(x)$ between $x = 0.03$ and $x = 0.04$ (Fig. \ref{F3})
hints on possible existence of an additional phase line on the $x - T$ phase diagram and calls for additional
studies of Cu$_x$TiSe$_2$ by other techniques, including scattering. Evolution of $\alpha_{100K}(x)$ (Fig.
\ref{F3}) is consistent with crossing the composition of the CDW transition at this temperature, whereas
$\alpha_{50K}(x)$ data suggest a gradual softening of the lattice on doping above $x = 0.04$. The fact that a
clear feature near $x \approx 0.03$ exists at all temperatures suggests that there may be a change in the nature
of the compound as $x$ increases through this value.

The origin of the step-like features in $\alpha(T)$ of Cu$_{0.08}$TiSe$_2$ and Cu$_{0.1}$TiSe$_2$ samples (Fig.
\ref{F2}(c)) is not clear at this point. It should be mentioned that TE measurements on polycrystalline samples
are potentially vulnerable to the morphology of the grains and grain boundaries and distribution of the grain
orientation in anisotropic materials, but an extrinsic mechanism causing a step-like behavior in $\alpha(T)$ of
single phase, polycrystalline, material is difficult to conceive of. This said, TE measurements on \textit{single
crystals} would be instrumental for understanding of these complex materials.

The observed features in temperature-dependent thermal expansion for different Cu concentrations (step-like
feature in $\alpha_{300K}$ \textit{vs} $x$ and change of sign of $\Delta \alpha$ at $T_{CDW}$) would be consistent
with a change in the sample and nature of CDW transition as we cross from low doping ($x \leq 0.03$) to
intermediate doping ($0.04 < x < 0.08$). For higher Cu-intercalation ($x \geq 0.08$) a new feature in $\alpha(T)$
appears, that is not associated with any line in the initial phase diagram \cite{mor06a} and may point to possible
structural distortion in highly Cu-intercalated samples at temperatures $\sim 160$ K or even related to the nearby
Cu-solubility limit ($x = 0.11 \pm 0.01$). \cite{mor06a}

\subsection{Superconductivity under pressure}

\begin{figure}[tbp]
\begin{center}
\includegraphics[angle=0,width=120mm]{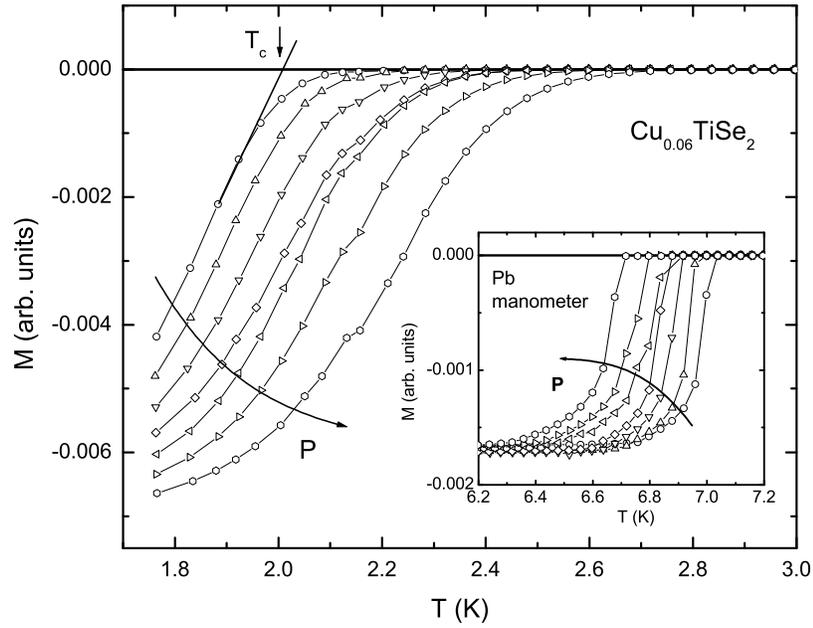}
\end{center}
\caption{Temperature dependent magnetization of Cu$_{0.06}$TiSe$_2$ measured in 25 Oe applied magnetic field under
pressures of 0.2, 1.6, 2.8, 4.2, 4.7, 6.1, and 8.7 kbar. $T_c$ was defined as an onset of magnetization. Inset
shows Pb superconducting transition measured. Arrow shows the direction of the pressure increase.} \label{F5}
\end{figure}

\begin{figure}[tbp]
\begin{center}
\includegraphics[angle=0,width=120mm]{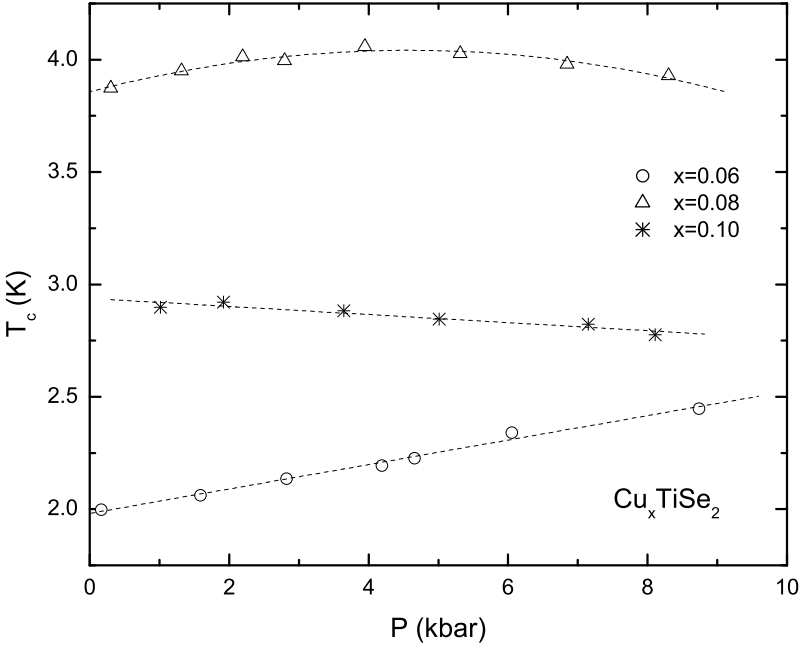}
\end{center}
\caption{Pressure dependent $T_c$ for Cu$_{0.06}$TiSe$_2$, Cu$_{0.08}$TiSe$_2$, and Cu$_{0.1}$TiSe$_2$ samples.}
\label{F6}
\end{figure}

\begin{figure}[tbp]
\begin{center}
\includegraphics[angle=0,width=120mm]{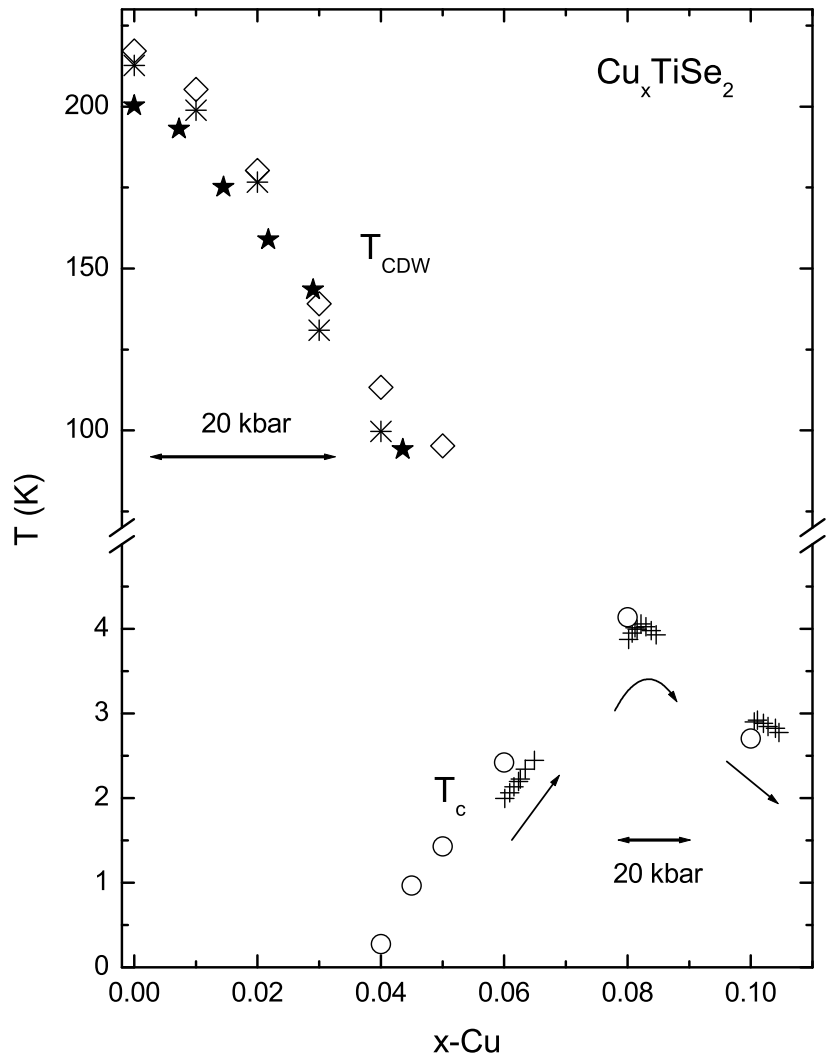}
\end{center}
\caption{Pressure dependent $T_c$ for Cu$_{0.06}$TiSe$_2$, Cu$_{0.08}$TiSe$_2$, and Cu$_{0.1}$TiSe$_2$ and
$T_{CDW}$ scaled on the $x - T$  phase diagram of Cu$_x$TiSe$_2$ \cite{mor06a}. Open symbols are taken from in
\cite{mor06a}, asterisks ($P = 0$, thermal expansion) and crosses (magnetization under pressure) are from this
work, stars are data from Ref. \cite{fri82a}. Horizontal bars show the (different) scaling factors between $x$ and
pressure for $T_{CDW}$ and $T_c$. Small differences in $T_c(P=0)$ between this work and published data are due to
differences in $T_c$ criteria used. Arrows indicate the direction of the pressure increase.} \label{F8}
\end{figure}

An example of magnetization measurements under pressure (for Cu$_{0.06}$TiSe$_2$) is shown in Fig. \ref{F5}. For
this sample $T_c$ increases under pressure without a clearly detectable change in superconducting transition width
of the sample or Pb manometer. Evolutions of the superconducting transition temperatures for three samples,
Cu$_{0.06}$TiSe$_2$, Cu$_{0.08}$TiSe$_2$, and Cu$_{0.1}$TiSe$_2$, as a function of pressure are shown in Fig.
\ref{F6}. Each of the samples behaves differently under increasing pressure: $T_c$ increases for $x = 0.06$,
decreases for $x = 0.1$, and has non-monotonic behavior with a broad maximum at around 4 kbar for $x = 0.08$. It
is noteworthy that the effect of pressure on $T_c$ of Cu$_x$TiSe$_2$ is rather small, $dT_c/dP \approx 0.054$
K/kbar for Cu$_{0.06}$TiSe$_2$ and $dT_c/dP \approx -0.018$ K/kbar for Cu$_{0.1}$TiSe$_2$. These differences are
not surprising if compared with the $T_c$ vs. $x$ behavior at ambient pressure reported in \cite{mor06a}. The
pressure data for the three samples can be approximately scaled with the same scaling factor ($x/P \approx 5.6
\cdot 10^{-4}$ kbar$^{-1}$) onto the superconducting "bubble" of the ambient pressure $x - T$ phase diagram (Fig.
\ref{F8}). So, apparently, increase of pressure and Cu intercalation have a similar effect on the
superconductivity of Cu$_x$TiSe$_2$. Although such a scaling is noteworthy as an empirical observation, it has to
be pointed out that both lattice parameters of Cu$_x$TiSe$_2$ \textit{increase} with Cu- intercalation
\cite{mor06a}. This rules out the unit cell volume or any lattice parameter alone to be a structural control
parameter for the observed scaling of $T_c$.

It seems enticing to check if the same scaling can be applied to the CDW transition. Fig. \ref{F8} shows that
change of $T_{CDW}$ of the pure TiSe$_2$ under pressure \cite{fri82a} can be scaled to the $T_{CDW}(x)$ behavior
reported in \cite{mor06a} as well, however the scaling factors for $T_{CDW}$ and $T_c$ differ by almost factor of
3, $1.5 \cdot 10^{-3} x/$kbar for $T_{CDW}$ \textit{vs} $5.6 \cdot 10^{-4} x/$kbar for $T_c$ (Fig. \ref{F8}). If
there would be the same one-to-one correspondence between Cu-intercalation and pressure in the $0 \leq x \leq 0.1$
Cu-concentrations range, we would expect to observe significantly (by almost factor of 3) higher pressure response
of $T_c$. This is consistent with pressure and Cu-doping both affecting the density of states and degree of
nesting in systematic and monotonic ways but by different mechanisms.

\section{Summary}
Thermal expansion measurements on polycrystalline Cu$_x$TiSe$_2$ samples confirmed the suppression of $T_{CDW}$ by
Cu-intercalation and suggested that the $x - T$ phase diagram may be more complex that in the original publication
\cite{mor06a}. These data raise the possibility that as Cu is added there is a change in the nature of the
compound and perhaps of the CDW transition for $x \geq 0.03$. The pressure data for Cu$_x$TiSe$_2$ samples ($x =
0.06, 0.08, 0.1$) can be approximately scaled with the same scaling factor on the superconducting "bubble" of the
ambient pressure $x - T$ phase diagram, however scaling of $T_{CDW}(P)$ data for pure TiSe$_2$ to the same phase
diagram will require a significantly different scaling factor.

Both sets of measurements suggest that the mechanism of how the superconducting state emerges from the CDW state
in Cu$_x$TiSe$_2$ and the relevant control parameter for this evolution of the ground state remains unclear and
additional experiments are required for a consistent physical picture.

\ack

Ames Laboratory is operated for the U. S. Department of Energy by Iowa State University under Contracts No.
W-7405-Eng.-82 and No. DE-AC02-07CH11358. Work at Ames Laboratory was supported by the director for Energy
Research, Office of Basic Energy Sciences. Work in Princeton was supported by the Department of Energy, Solid
State Chemistry Program, grant DE-FG02-98-ER45706. GMS is supported by the National Science Foundation under
DMR-0305397. SLB would like to thank Milton Torikachvili for useful discussions on techniques of samples' shaping.

\clearpage

\clearpage

\clearpage

\clearpage

\clearpage

\clearpage

\end{document}